\documentstyle[prd,aps,epsfig]{revtex}
\newcommand{\omicron}o

\begin{document}
\draft
%%%--uncomment for twocolumn
\twocolumn[\hsize\textwidth\columnwidth\hsize\csname
@twocolumnfalse\endcsname
%%%--
\title{DYNAMICS AND INTEGRABILITY PROPERTY\\
OF THE CHIRAL STRING MODEL}
\author{Brandon Carter and Patrick Peter}
\address{D\'epartement d'Astrophysique Relativiste et de Cosmologie,\\
Observatoire de Paris-Meudon, UMR 8629, CNRS, 92195 Meudon, France.}

\date{5 May 1999, preprint no DARC/99-09}

\maketitle

\begin{abstract} The effect of fermionic string conductivity by purely
right (or purely  left) moving ``zero modes''  is shown to be governed
by   a simple Lagrangian   characterising  a certain  ``chiral'' (null
current carrying)  string model  whose dynamical  equations  of motion
turn out to be explicitly integrable in a flat spacetime background.
\end{abstract}

\vskip2pc
%%%--uncomment for twocolumn
]
%%%--

\section{Introduction.}

In his epoch making paper entitled ``Superconducting Strings'', Witten
introduced  two essentially   different  processes   that  can produce
conserved currents  on cosmic strings.  The first, for  which the term
``superconducting'' is most appropriate,  is based on the formation on
the string of  a bosonic condensate  that is analogous  to the kind of
condensate  that is  familiar  in ordinary  laboratory superfluids and
superconductors. The second process   considered by Witten involves  a
purely fermionic  mechanism (of a  kind for  which -- since  no Cooper
type pairing  is  involved  --  the term ``superconducting''    is not
appropriate) based on ``zero modes''.

In  order  to describe   the macroscopic  effect of   the (classically
conserved) currents   arising  from  such  processes, whether   of the
bosonic ``superconducting'' kind  or the fermionic ``zero mode'' kind,
Witten~\cite{Witten} proposed  the use  of  a string model  that  is a
simple --    in a certain  sense linearised   -- generalisation of the
Nambu-Goto model  for  an  ``ordinary''  non conducting  string.  With
respect to internal   coordinates  $\sigma^i$, in  a background   with
coordinates $x^\mu$   and spacetime   metric $g_{\mu\nu}$,   a generic
conservative string  model will be characterised by   an action of the
form
\begin{equation} {\cal I}=\int {\cal L}\, d^{(2)}\!{\cal S}\, ,
\hskip 1 cm  d^{(2)}\!{\cal S}= \Vert\gamma\Vert^{1/2} d\sigma^{_0}
d\sigma^{_1} \,  , \label{0} \end{equation}  where ${\cal L}$  is a function of the
relevant internal variables, if any, and  where $\vert \gamma\vert$ is
the determinant of the induced metric given by
\begin{equation} \gamma_{ij}=g_{\mu\nu} x^\mu_{\, ,i} x^\nu_{\, ,j}\label{1a}\end{equation} 
using a comma for partial differentiation. 

In  the  Nambu Goto case   there  are no internal  variables,  so  the
Lagrangian will just be a constant, having the form
\begin{equation}{\cal L}=-m^2   \label{2}   \end{equation}
for a fixed parameter $m$ which -- with respect to units such that the
Dirac-Planck constant $\hbar$ and the speed of light $c$ are set equal
to unity -- will have  the dimensions  of a  mass.  (In the case of  a
``ordinary'' cosmic string model, representing a  vortex defect of the
vacuum,  $m$ would be  expected~\cite{Kibble}  to be  of  the order of
magnitude of the relevant Higgs mass scale).

In the Witten model, there is  an internal scalar phase field variable
$\varphi$ in terms of which the action takes the less trivial form
\begin{equation}{\cal L}=-m^2-{_1\over^2}\kappa_{_0}\gamma^{ij}\varphi_{|i}
\varphi_{|j} \, , \label{3} \end{equation}
for some positive  constant  $\kappa_{_0}$.   In  the presence of    a
Maxwellian gauge covector   field $A_\mu$, if  the  internal field has
charge coupling constant $e$, the  relevant gauge covariant derivative
will have   the form $\varphi_{|i}=\varphi_{\,   ,i}-e A_\mu x^\mu_{\,
,i}$.

Soon after its introduction, it  was pointed out  that for the purpose
of     treating       Witten's       bosonic     ``superconductivity''
mechanism~\cite{Witten}, his   simple   bosonic model  (\ref{3}) would
require generalisation~\cite{C89}  to  allow for non-linearity.   More
particularly,  it was shown  by one of us~\cite{Peter92} that Witten's
bosonic ``superconductivity''  mechanism    entailed supersonic wiggle
propagation,  whereas his  own proposed   string model (\ref{3})   was
characterised by subsonic  wiggle propagation. This discrepancy, which
is significant even for very weak currents, has recently been resolved
by replacing  Witten's naive model  (\ref{3}) by  a more appropriately
adapted  model~\cite{CP95} whose action  is non  linearly dependent on
the current amplitude.

In so far as  Witten's fermionic ``zero-mode''  conductivity mechanism
is concerned,  the question of the adequacy   or otherwise of Witten's
naive model  (\ref{3}) has not yet been  given much attention. Whereas
it is evident  that corrections  will  be needed  for cases  involving
strong   currents, what is  not yet   clear  is whether Witten's model
(\ref{3}) will be invalidated (as  in the bosonic case) by  non-linear
effects even for fermionic currents of very low amplitude.

There is however a special case -- the  subject of the present article
-- for which  it  seems clear  in advance that  there  is no  risk  of
invalidation by  non  linear effects,   because  the relevant  current
amplitude simply vanishes -- a condition that will be preserved by the
field  equations  of the model (\ref{3})   if $e=0$. The  condition of
having zero amplitude does not of course mean  that the current itself
has to vanish, but merely that it should be null in the sense of being
lightlike. Individual   fermionic    ``zero mode''   excitations   are
intrinsically  characterised by  the  condition of  being null in this
sense, but a generic   superposition of ``left'' and  ``right'' moving
excitations  will provide a    total  current  that  is  timelike   or
spacelike.  However, in  the special  ``chiral''  case for  which only
``left''  (or equivalently only   ``right'') moving ``zero modes'' are
excited, the total current will be null, i.e. it will satisfy
\begin{equation} \gamma^{ij}\varphi_{,i} \varphi_{,j} =0\, .\label{4}\end{equation}

The suggestion that such   a ``chiral'' current might arise  naturally
from the  presence of massive neutrinos in  Grand Unified Theories was
implicit  in  Witten's   original article~\cite{Witten}  [right-handed
neutrino current in    $SO(10)\to SU(5)\times \tilde   U(1)\to  SU(5)$
theory with the  last symmetry breaking  realized by means of  a Higgs
field in the {\bf 126} of SO(10)],  and the question has recently been
raised more   explicitly~\cite{PerkinsDavis97} by several  authors: in
particular  it  has   been  pointed   out~\cite{Davis2Trodden}    that
``chiral'' (purely left  or purely right moving)  zero modes are to be
expected in strings with half broken supersymmetry  in the presence of
a   Fayet  Iliopoulos term.  The    technical requirements for  having
lightlike  currents  flowing along  the  strings due   to fermions are
discussed in the appendix.

The likelihood that null currents can  arise naturally from ``chiral''
zero   modes has potentially   important cosmological implications, in
view   of  the consequent possiblity of   formation  of chiral vortons
--which can be expected to be more stable  than vortons of other kinds
-- a perspective that has provided the main motivation for the present
investigation.

Quite appart from the consideration that it will be applicable exactly
in such cases,  the chirality condition  (\ref{4}) has also frequently
been  advocated  for  use  as   an  approximation  even  for   bosonic
currents. The first occasion this was done  was in the seminal article
~\cite{DavisShellard89}  in  which the question   of vorton states was
originally  raised  by  Davis  and Shellard,    and  the approximation
(\ref{4}) has in fact been  systematically employed in the more recent
work~\cite{MartinsShellard98}.    In the particular  case  of circular
string loops the solution  are characterised~\cite{CPG97} by a certain
Bernouilli  ratio, $b$, that  would need to  be close to unity for the
chiral approximation to be tenable. However even when such a condition
is satisfied the use~\cite{MartinsShellard98} of such an approximation
is dangerous, since   it artificially supresses   potentially unstable
degrees of  freedom  that  are  present when  the   generic non linear
model~\cite{CP95} is used.

Although its  use as  an  approximation  for the treatment  of  string
currents  that are   actually    spacelike  or  timelike is     rather
questionable, there can be  no such doubts  about the validity of  the
chiral    string model in   cases  when    the current is   physically
constrained to  be  null, as  will be  the  case when  it  arises from
fermionic zero modes of   the   chiral type discussed above.    Chiral
solutions are obtainable as solutions for string models of the generic
non-linear      elastic     type~\cite{C89,CP95}     including     the
Witten~\cite{Witten}  limit  case  (\ref{3})    (in the   absence   of
electromagnetic  coupling,  i.e. for  $e=0$) subject   to  the nullity
constraint (\ref{4}).   One of the first  points to be emphasised here
is that the chiral string model can also conveniently be characterised
directly  in its own  right by a Lagrangian  of  its own, which can be
taken to have the form
\begin{equation} {\cal L}=-m^2-{_1\over^2}\psi^2\gamma^{ij}\varphi_{,i} \varphi_{,j}
\, . \label{5} \end{equation}
This chiral string  Lagrangian  is obtained by replacing  the positive
constant $\kappa_{_0}$ in  the Witten~\cite{Witten} model (\ref{3}) by
the square of an auxiliary  Lagrange multiplier field $\psi$.  For the
purpose of applying the  variation principle, the  auxiliary amplitude
$\psi$ and the phase  scalar $\varphi$ are to  be treated on  the same
footing  as independent   internal  field  variables  on  the   string
worldsheet, whose   spacetime location  is  of   course  also  to   be
considered to be freely variable.

\section{Dynamical Equations of the Chiral String Model.}

It is evident  that  the requirement of   invariance of  the  integral
(\ref{0}) with respect to free variations of $\psi$ in (\ref{5}) leads
back immediately to   the nullity condition (\ref{4}).  In particular,
since the worldsheet  is only 2-dimensional,  this has the  well known
corollary that the field $\phi$ will be harmonic, i.e.
\begin{equation} \varphi^{;i}_{\ ;i}=0\, ,\label{6}\end{equation}
using a semicolon  to indicate covariant  differentiation with respect
to    the   worldsheet   metric    $\gamma_{ij}$.     Furthermore,  by
differentiation of  (\ref{4}), it can  be seen the phase gradient must
satisfy the geodicity condition
\begin{equation} \varphi_{;ij}\varphi^{;j}=0\, .\label{7}\end{equation} 

The only other internal field equation on  the world sheet, namely the
one  obtained from   the  requirement of   invariance of  the integral
(\ref{0}) with respect to free variations of $\phi$ in (\ref{5}), will
evidently take the form
\begin{equation}  \big(\psi^2\varphi^{;i}\big)_{;i}=0\, .\label{8}\end{equation}
In view of (\ref{6}), the latter reduces simply to the condition
\begin{equation} \varphi^{;i}\psi_{,i}=0\, ,\label{9}\end{equation}
to the effect that $\psi$ is constant along the world sheet null lines
on which  $\varphi$ is constant, or in  other  words that $\psi$  is a
function only of $\varphi$.

For the purpose   of dealing with   the extrinsic equations  of motion
governing the evolution of the worldsheet it  is convenient to work in
terms of tensors specified  with  respect to the  background spacetime
coordinates     $x^\mu$,    starting with   the    fundamental  tensor
$\gamma^{\mu\nu}$ that    is   defined~\cite{C89}  as  the   spacetime
projection  of the contravariant inverse  $\gamma^{ij}$ of the induced
metric, i.e.
\begin{equation}\gamma^{\mu\nu}=\gamma^{ij} x^\mu_{\, ,i} x^\nu_{\, ,j} \,
,\label{10}\end{equation} (whose mixed form $\gamma_\mu^{\, \nu}$ is the tangential
projection tensor) and the tangent vector
\begin{equation}  \nu^\mu=\psi \varphi^{;i} x^\mu_{\, ,i}\, .\label{11}\end{equation}
The  latter   will  be   equivalently expressible   in  terms  of  the
tangentially projected covariant differentiation operator
\begin{equation} \overline\nabla_{\!\mu}=\gamma_\mu^{\,\nu}\nabla_{\!\nu}\,
,\label{12}\end{equation}
(where $\nabla_{\!\nu}$    is  the    usual   operator  of   covariant
differentiation with respect to $g_{\mu\nu}$) in the form
\begin{equation} \nu^\mu=\psi \overline\nabla{^\mu}\varphi\, .\label{13}\end{equation}

In terms of this worldsheet  tangential vector $n^\mu$, the Lagrangian
(\ref{5}) of the chiral string model can be rewritten in the form
\begin{equation} {\cal L}=-m^2-{_1\over^2}\nu^\mu \nu_\mu\, , \label{14} \end{equation}
and it  can  be seen that    the ensuing internal  field equations  as
obtained above will be expressible as the nullity condition
\begin{equation}  \nu^\mu \nu_\mu=0\, ,\label{15}\end{equation}
and the worldsheet current conservation law
\begin{equation} \overline \nabla_{\!\mu} \nu^\mu=0\, .\label{16}\end{equation}

The  standard  procedure~\cite{C95}    for obtaining  the    extrinsic
equations of motion starts from the evaluation of the relevant surface
stress  momentum energy density  tensor   as  defined by the   general
formula
\begin{equation} \overline T{^{\mu\nu}}=2{\partial{\cal L}\over\partial g_{\mu\nu}}
+ {\cal L}\gamma^{\mu\nu}\, , \label{17}\end{equation}
which in the present application (\ref{14}) simply gives
\begin{equation} \overline T{^{\mu\nu}}=\nu^\mu \nu^\nu - m^2\gamma^{\mu\nu} 
\, ,\label{18}\end{equation}
when the on-shell condition (\ref{15}) is satisfied.  In terms of this
tensor, the   dynamical condition for  the  invariance  of  the action
(\ref{0}) with  respect  to arbitrary  infinitesimal displacements  is
equivalent to the condition of vanishing external force
\begin{equation} f^\mu=0 \, ,\label{19}\end{equation}
where the force density is given by an expression of the usual form
\begin{equation} f^\mu= \overline\nabla_{\!\nu}\overline T{^{\nu\mu}}\, .
\label{19a}\end{equation}
as the condition for the invariance of the action (\ref{0}) with
respect to arbitrary infinitesimal displacements.

As usual the tangential projection of (\ref{19})  will be satisfied as
a Noether identity, i.e. we shall automatically have
\begin{equation} \gamma^\rho_{\ \mu} f^\mu=0\, ,\label{20}\end{equation}
when   the relevant internal equations,  in   this case (\ref{15}) and
(\ref{16}),  are  satisfied. This means   that   the evolution of  the
worldsheet  will    be given just  by  the   orthogonal  projection of
(\ref{16}), i.e. its contraction
\begin{equation} \perp^{\!\rho}_{\,\mu} f^\mu=0\, ,\label{21}\end{equation}
where  the  orthogonal  projection  tensor $\perp^{\!\rho}_{\,\mu}$ is
defined as the complement of the fundamental tensor $\gamma^\rho_{\
\mu}$, i.e.
\begin{equation} \perp^{\!\rho}_{\,\mu}= g^\rho_{\ \mu}-\gamma^\rho_{\ \mu}\,
.\label{22}\end{equation}   By  the usual integration      by parts operation  this
orthogonally projected force density  on the worldsheet is obtained in
the standard form
\begin{equation}  \perp^{\!\rho}_{\,\mu} f^\mu=
\overline T{^{\mu\nu}} K_{\mu\nu}{^\rho}\, ,\label{23}\end{equation}
where the $K_{\mu\nu}{^\rho}$ is   the  second fundamental tensor   as
defined~\cite{C89} in terms of tangential differentiation of the first
fundamental tensor $\gamma^\rho_{\ \sigma}$ by
\begin{equation} K_{\mu\nu}{^\rho} = \gamma^\sigma_{\ \nu}\overline\nabla_{\!\mu}
\gamma^\rho_{\ \sigma} \, .\label{24}\end{equation}
It evidently follows that in the  particular case of the chiral model,
as  characterised   by (\ref{18}),   the extrinsic  dynamical equation
(\ref{21}) will take the form
\begin{equation} m^2 K^\rho=\nu^\mu\nu^\nu  K_{\mu\nu}{^\rho}\, ,\label{25}\end{equation}
where  the  extrinsic curvature vector  $K^\rho$ is  the  trace of the
second fundamental tensor, i.e.
\begin{equation} K^\rho=K_\mu^{\ \mu\rho}\, , \label{26}\end{equation}
(whose  vanishing expresses the equation of  motion for the Nambu Goto
model, as obtained  in the  limit  for which  the null current  vector
$\nu^\mu$  vanishes so  that the  right  hand side of (\ref{25}) drops
out).

\section{Characteristic formulation of the dynamical equations.}

In  order to  proceed it is   convenient to introduce a conventionally
normalised null   tangent vector diad on   the worldsheet,  taking one
member  to  be the conserved null   current  vector $\nu^\mu$ that has
already been  introduced, which can   without loss of generallity  (if
necessary by adjusting the sign convention for $\varphi$ or $\psi$) be
assumed to be oriented towards  the future.  Subject to the convention
that  it too  should be oriented  towards  the  future  the other null
tangent vector, $\omicron^\mu$ say, will be characterised by
\begin{equation} \omicron^\mu\omicron_\mu=0\, \hskip 1 cm
\omicron^\mu\nu_\mu= -1\, ,\label{27}\end{equation}
so that the fundamental tensor will take the form
\begin{equation} \gamma^{\mu\rho}=-2\nu^{(\mu}\omicron^{\rho)}\, ,\label{28}\end{equation}
using round brackets to denote  index symmetrisation.  The  worldsheet
stress momentum energy density tensor (\ref{18})  can then be given an
expression of the analogous form
\begin{equation} \overline T{^{\mu\rho}}=\nu^{(\mu}\beta^{\rho)}\, ,\label{29}\end{equation}
in which $\beta^\nu$ is a future directed timelike
worldsheet tangent vector given by
\begin{equation} \beta^\mu=\nu^\mu+2m^2\omicron^\mu\, .\label{30}\end{equation}

In  view of  the general   theorem~\cite{C95} to  the effect  that the
characteristic covectors $\chi_\mu$  (i.e. gradients  of scalars  that
are   constant   along   admissible  directions   of   propagation  of
infinitesimal discontinuity) of  the extrinsic motion are specified as
solutions of the quadratic characteristic equation
\begin{equation} \overline T{^{\mu\rho}}\chi_\mu\chi_\rho=0\, ,\label{31}\end{equation}
it is evident  from (\ref{29}) that the  null vector $\nu^\mu$ defined
by (\ref{11})   and  the  timelike   vector  $\beta^\mu$  defined   by
(\ref{30})  lie  along  the corresponding  bicharacteristic directions
(i.e.  admissible  directions    of   propagation  of    infinitesimal
discontinuity) of the extrinsic motion. It is also evident that as far
as the  internal dynamical   equations (\ref{15}) and   (\ref{16}) are
concerned there  is only  a  single bicharacteristic  direction, which
coincides with the extrinsic  bicharacteristic direction  specified by
$\nu^\mu$.

The preceeding conclusion means that the dynamical behaviour of chiral
string model  will  be much simpler  than  that of  a generic  elastic
string model~\cite{C95}  or that of  the Witten model~\cite{Witten} in
particular, for which there  are two independent internal (sonic type)
bicharacteristic directions, neither of which coincides with either of
the two extrinsic bicharacteristic  directions. The case of the chiral
string  model  is intermediate between that   of the special transonic
string model~\cite{C95a} for  which there are two independent internal
bicharacteristic directions   which   exactly coincide  with   the two
extrinsic bicharacteristic  directions,  and that of  the simple Nambu
Goto model, which has  no internal degrees of freedom,  so that it has
no internal bicharacteristic directions at all.

Following the    example   of  the  transonic  case~\cite{C95a},   the
coincidence   of the only  internal  bicharacteristic direction of the
chiral string   model  with  one  of  its   extrinsic bicharacteristic
directions can  be  exploited   for  the purpose  of  converting   its
dynamical   equations  to   a particularly   convenient characteristic
form.  To start with, as is  possible for any string model~\cite{C95},
the characteristic expression (\ref{29}) for the surface stress tensor
can be substituted  into the generic  force equation (\ref{19a})  from
which, using the worldsheet integrability condition
\begin{equation} \perp^{\!\rho}_{\,\mu}\big(\beta^\nu\nabla_{\!\nu}\nu^\mu
-\nu^\nu\nabla_{\!\nu}\beta^\mu\big)=0 \, ,\label{32}\end{equation} one obtains the
extrinsic (orthogonally projected) part  of the force equation in  the
convenient form
\begin{equation} \perp^{\!\rho}_{\,\mu} f^\mu= 
\perp^{\!\rho}_{\,\mu}\nu^\nu\nabla_{\!\nu}\beta^\mu\, .\label{33}\end{equation}

In the particular  case of  the chiral  model,  the internal dynamical
equation          (\ref{15})       evidently     implies      $\nu^\mu
\overline\nabla_{\!\rho}\nu_\mu=0$, from which by (\ref{28}) it can be
seen that the  other internal  dynamical  equation (\ref{16}) will  be
expressible in either of what, by (\ref{27}), are the equivalent forms
\begin{equation} \omicron^\mu\nu^\rho\nabla_{\!\rho}\nu_\mu=0 \hskip 1 cm 
\nu^\mu \nu^\rho\nabla_{\!\rho}\omicron_\mu=0\, .\label{34}\end{equation}
By  further use of (\ref{15}),  (\ref{27})  and (\ref{28}), it can  be
seen that the preceeding relations are equivalent to
\begin{equation} \gamma^\rho_{\ \mu}\nu^\nu\nabla_{\!\nu} \nu^\mu=0\, ,\hskip 1 cm
 \gamma^\rho_{\        \mu}\nu^\nu\nabla_{\!\nu}\omicron^\mu=0      \,
 ,\label{35}\end{equation}  which    by  (\ref{30})   will  also  be   equivalently
 expressible as
\begin{equation} \gamma^\rho_{\ \mu}\nu^\nu\nabla_{\!\nu} \beta^\mu=0\, .\label{36}\end{equation}

Using the   analogous formula(\ref{33}) it can now   be seen that this
internal dynamical equation  (\ref{36})  can be amalgamated  with  the
extrinsic  dynamical equation (\ref{21})  to give a combined dynamical
equation of the simple characteristic form
\begin{equation} \nu^\nu\nabla_{\!\nu}\beta^\mu=0\, , \label{37}\end{equation}
which effectively   states  that the   timelike  characteristic vector
$\beta^\mu$ is subject to transport by  the null characteristic vector
$\nu^\mu$. This      transport  equation  evidently     preserves  the
normalisation conditions
\begin{equation} \beta^\mu \nu_\mu=-2m^2\, ,\hskip 1 cm \beta^\mu \beta_\mu=-4m^2
\, ,\label{38}\end{equation}
that follow from the defining relation (\ref{30}).

As in the familiar example  of the Nambu Goto case  (for which both of
the extrinsic  characteristic directions  are  null) it  is useful for
many  purposes   to  use coordinates   that  are constant   along  the
respective    characteristic   directions.  As    far   as  the   null
characteristic direction is concerned, the scalar $\varphi$ introduced
above    can  evidently serve   for   this purpose.    With respect to
worldsheet    coordinates   consisting    of  $\varphi$   and  another
characteristic  coordinate, $\sigma$ say,  chosen to be constant along
the timelike characteristic direction specified by $\beta^\mu$, it can
be seen from  (\ref{11}) and  (\ref{38})  that, whatever the   precise
convention used  to fix the normalisation   of $\varphi$ and $\sigma$,
the timelike characteristic vector will be given by
\begin{equation} \beta^\mu= -2{m^2\over\psi}{\partial x^\mu\over\partial\varphi}
\, .\label{39}\end{equation} 

\section{Integrability in a flat background.}

The preceeding equations   are applicable  in an  arbitrarily   curved
cosmological background spacetime. Let us  now restrict our  attention
to  the case  of a  flat  background, and   more specifically let   us
restrict  the  coordinates   to   be  of  ordinary  Minkowski    type,
$x^\mu\leftrightarrow \{x^{_0},  x^a\}$, $a=1,2,3$  so that the metric
is given by
\begin{equation} g_{\mu\nu}\, dx^\mu dx^\mu=(dx^{_0})^2-
\delta_{ab}\,dx^a dx^b \label{40}\end{equation}
where   $\delta_{ab}$   is the    Kronecker  unit  matrix.   Since the
Christoffel connection  will  vanish in such  a reference  system, the
characteristic dynamical equation (\ref{37}) will reduce to the simple
form
\begin{equation} {\partial\over\partial \sigma}\beta^\mu=0\, .\label{41}\end{equation}
Since by (\ref{9}) we know that $\psi$ depends only on $\varphi$,
it can be seen from (\ref{39}) that (\ref{41}) is equivalent to
\begin{equation} {\partial^2 x^\mu\over \partial \sigma\partial\varphi}=0\, , 
\label{42}\end{equation}
whose general solution  is given as  the  sum of a  pair of generating
curves $\zeta^{\mu}\{\sigma\}$, $\xi^{\mu}\{\varphi\}$ parametrised by
the  characteristic coordinates  $\sigma$  and $\varphi$ respectively,
i.e.
\begin{equation} x^\mu\{\sigma,\varphi\}=
\zeta^{\mu}\{\sigma\}+\xi^{\mu}\{\varphi\}\, .\label{43}\end{equation}

In order for   such a solution to  represent  a physically  admissible
configuration for the  chiral string model,  the generating curves can
not be chosen in an entirely arbitrary manner: in order to satisfy the
condition (\ref{15})  that $\nu^\rho$ should  be null,  the tangent to
the first generating curve $\zeta^{\mu}\{\sigma\}$ must be null, while
in order for $\beta^\mu$ to  be timelike  the  same must apply to  the
tangent  to the second   generating curve $\xi\{\varphi\}$, i.e. using
the notation
\begin{equation} \zeta{^{\prime\mu}}={d \zeta^\mu\over d\sigma}\, ,\hskip 1 cm
\dot \xi{^\mu}={d \xi^\mu\over d\varphi}\, ,\label{44}\end{equation}
we must construct the generating curves  so as to satisfy the equality
and inequality
\begin{equation} \zeta{^{\prime\mu}}\zeta^\prime_{\mu}= 0\, ,\hskip 1 cm 
\dot\xi{^\mu}\dot\xi{^\mu}< 0 \, .\label{45}\end{equation}
The chiral string worldsheet solution obtained in this way can be seen
to   differ  from   its well  known     analogue for  the Nambu   Goto
case~\cite{KibbleTurok} only by the  condition that in the Nambu  Goto
case both generating curves are restricted to be  null, whereas in the
chiral  case this  restriction  applies   to only   one of  them.  The
transonic string model is  also   solved by  an   ansatz of the   form
(\ref{43}),  but  in this  case~\cite{C95a}  neither of the generating
curves is restricted to be null.

As in the  well known Goto  Nambu example, it  will be  convenient for
many purposes to    use  a background  time   parametrisation  for the
generating curves, i.e. to take
\begin{equation} \sigma =\zeta^{_0}\, , \hskip 1 cm \varphi=\xi^{_0}\, ,\label{46}\end{equation}
in which case (\ref{45}) will be expressible as the condition that the
space vector $\zeta{^{\prime   a}}$ must lie  on  the ``Kibble Turok''
unit sphere,  while $\dot  \xi{^a}$ must  lie inside the  unit sphere,
i.e.
\begin{equation} \delta_{ab}\,\zeta{^{\prime a}} \zeta{^{\prime b}} =1\,  ,\hskip 1
cm \delta_{ab}\, \dot \xi{^a} \dot \xi{^b} < 1 \, .\label{47}\end{equation}

Whatever   parametrisation  convention is used,  it   can be seen from
(\ref{39})  that the  null  and   timelike generators $\nu^\mu$    and
$\beta^\mu$ will be respectively given by
\begin{equation} \nu^\mu=\psi\,\big(\dot\xi{^\nu}\zeta^\prime_\nu\big)^{-1}
\zeta{^{\prime\mu}} \, ,\hskip 1 cm \beta^\mu= 
-2{m^2\over\psi}\,\dot\xi{^\mu}\, ,\label{48}\end{equation}  in which, in  order to
satisfy the normalisation   condition (\ref{38}), the auxiliary  field
amplitude $\psi$ will be given by
\begin{equation} \psi^2=-m^2\, \dot\xi{^\mu} \dot\xi_\mu\, .\label{49}\end{equation}
The corresponding explicit expression for the surface stress tensor of
the chiral string can thus be seen from (\ref{29}) to be given by
\begin{equation} T{^{\mu\rho}}= -2 m^2\,\big(\dot\xi{^\nu} \zeta^\prime_\nu\big)^{-1}
\zeta{^{\prime(\mu}} \dot\xi{^{\rho)}}\, .\label{50}\end{equation}

\section{Conclusions.}

Having   just  one internal  degree   of  freedom --  corresponding to
longitudinal  modes  propagating  at the speed  of  light  in just one
(forward but not backward) direction -- the chiral string model can be
described as  being intermediate between the  Nambu Goto  string model
which has no  internal  degrees of  freedom  at all, and  the  generic
elastic  models~\cite{CP95}  (of which the  Witten~\cite{Witten} model
(\ref{3}) was the original prototype) which  have two internal degrees
of freedom corresponding to  longitudinal propagating in both left and
right directions, with  a  ``sound'' speed  that  is equal to that  of
light in the  case of the   naive Witten model~\cite{Witten} but  that
will  be slower for  more  realistic  models~\cite{CP95}. There  is  a
general theorem ~\cite{CarterMartin93}   to  the  effect  that  vorton
equilibrium states will   be classically stable  whenever the relevant
propagation speed  of extrinsic ``wiggles''  is subsonic or transonic.
The condition of subsonicity is  always satisfied for the Witten model
(\ref{3}), which is precisely the reason  why this model is physically
unnaceptable    for      the    representation  of    bosonic   string
``superconductivity'' in which, at least for weak currents, the wiggle
propagation speed has  been predicted~\cite{Peter92} to be supersonic,
with the implication that  the corresponding vorton  stability problem
will be rather complicated~\cite{CarterMartin93,MartinPeter95}.  It is
to  be  similarly remarked   that the  condition  of  transonicity  is
automatically  satisfied for  the chiral string  model (\ref{5}), with
the implication that this model too is potentially misleading, in that
it       can  give  a      false    impression    of  stability,  when
used~\cite{MartinsShellard98}       as  an  approximation  for     the
representation of string currents  of the bosonic  ``superconducting''
type.  This caveat needs to be particularly emphasized  in view of the
very convenient integrability  property established  in the preceeding
section, which reinforces   the temptation to  use  the  chiral string
model for applications beyond its range of legitimate validity.

Although     potentially   misleading        when used      as      an
approximation~\cite{MartinsShellard98} in the bosonic case, the use of
the  chiral model (\ref{5})      --  and the implication    that   the
corresponding vorton  states will always be  classically  stable -- is
physically justifiable  as a valid description in  cases for which the
current under consideration arises from fermionic zero modes of chiral
type  such     as are  expected   to arise    in   some grand  unified
theories~\cite{Witten}   or certain  scenarios  involving   incomplete
supersymmetry breaking~\cite{Davis2Trodden}.

\section*{Appendix: elimination of fermionic current anomalies.}

The purpose of this appendix is to show how  the possibility of having
conserved -- anomaly free -- chiral currents  on the string depends on
the  requirement that  they  consist of  axial fermions   that (as was
assumed in  the preceding work) are  not coupled to  any long range --
electromagnetic or   other  -- gauge field.  We   start   by using the
Nielsen-Olesen string solution~\cite{NO}  to check that in the absence
of long range field coupling  the two-dimensional theory in the string
background will be anomaly  free even though  the background itself is
nontrivial.  We then go on to consider a prototypical four-dimensional
theory admitting a  Nielsen-Olesen string solution that contains  also
fermionic carriers coupled  to both  the strings  and  some extra long
range -- electromagnetic type -- gauge fields.

\begin{figure}
\centering
\epsfig{figure=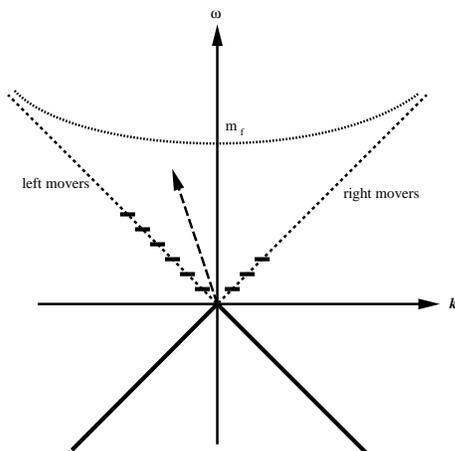, width=6cm}
\caption{Construction of timelike current from zero mode solutions
of the Dirac equation in a cosmic string background for the
generic charge coupled case, in which there are zero modes 
propagating in both left and right null directions.
\label{fig1}}
\end{figure}

The essential point  is that unless they  are entirely decoupled  from
the axial fermions constituting the string  current, such fields would
give   rise to   anomalies    whose cancellation  would  require   the
introduction of  other fermions, at   the cost of  spoiling the chiral
nullity  condition~(\ref{4}),   in the  manner  illustrated  in figure
\ref{fig1}.

The four-dimensional model we shall have  in mind consists of a gauged
Abelian Higgs model characterised by a Lagrangian containing a bosonic
contribution ${\cal  L}_{_{A.H}}$ that  is coupled  via  a Yukawa term
${\cal L}_{_Y}$ to  a set of fermionic  fields denoted by $\Psi$  in a
total of the form $ {\cal L}$ $={\cal L}_{_{A.H}} ({\cal H},B_\mu ) +$
$   {\cal     L}_{_Y}({\cal   H},\Psi)     +$   $\sum    \overline\Psi
\mbox{\boldmath$\gamma$} ^\mu  D_\mu \Psi$, in   which the final gauge
interaction term  involves a sum  running  over all the  fermions that
couple  to  the strings (i.e. all  those  having non trivial zero mode
solutions  in a string background).  For  each individual fermion, the
couplings  with the gauge fields  have  the form $  D_\mu  \Psi $ $= (
\nabla _\mu - i e Q A_\mu -i g R B_\mu)\Psi\ , $  in which $Q$ and $R$
are  the (hyper)charges of  the fermion in  question, and in which $e$
and   $g$ the coupling   constants, where  $A_\mu$ is   the long range
electromagnetic type gauge vector,  while $B_\mu$ is the gauge  vector
associated with the symmetry whose spontaneous breaking is responsible
for the existence  of the strings.  In  order for the string solutions
of  ${\cal L}_{_{A.H.}}$ to  admit zero modes, the fermionic couplings
with   $B_\mu$   must  be  different   for the   left-handed   and the
right-handed  ones~\cite{Witten},  thereby potentially generating  the
so-called $QQR$ and $QRR$ anomalous terms.

\begin{figure}
\centering
\epsfig{figure=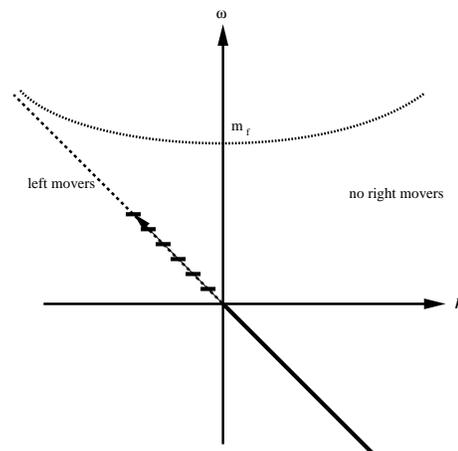, width=6 cm}
\caption{Construction of null current from zero mode solutions
of the Dirac  equation in a  cosmic string  background  for the chiral
case (requiring the  absence of charge  coupling) for which there  are
zero modes  propagating in a  single (in this illustration, left) null
direction.
\label{fig2}}
\end{figure}

The possibility of having purely chiral,  let us say leftwards moving,
currents in the  string depends  on the  condition  that the  relevant
fermionic   fields should  be of the   corresponding,  let us say left
handed, chiral type in the four  dimensional sense, since their Yukawa
coupling to the string-forming  Higgs field ${\cal  H}$ will give rise
to zero modes propagating in  only one direction, with  anti-particles
propagating in  the same direction~\cite{JackiwRossi},  as illustrated
in figure \ref{fig2}.

Since a  chirally  opposite, i.e. right   handed,  fermion field would
similarly couple with the Higgs  field conjugate ${\cal H}^\star$  (so
that the  vortex appears as   an anti-vortex) the  corresponding modes
would propagate in   the  opposite  direction,  i.e.    they  would be
rightwards  movers. In the generic  case for which  both kinds of mode
are present, their simultaneous occupation gives rise to currents that
need no longer be null but can be timelike, as illustrated in figure
\ref{fig1}.

Provisionally setting  aside   the possibility  of   coupling with  an
external field  $A_\mu$,   by  setting  $e=0$, let  us    consider the
possibility  of   anomalies  in  the string  itself,  considered  as a
two-dimensional  worldsheet admitting   zero  modes of   the  relevant
fermionic model   in a given   gauge background.   For  such a  model,
anomalies might arise  at the one-loop  level because  of the coupling
between the fermions and the string-forming  gauge field $B_\mu$ whose
presence implies a nonvanishing background $G_{\mu\nu} = \partial _\mu
B_\nu  -  \partial_\nu B_\mu$.  In  such a   theory  one finds that in
general~\cite{67} the  axial  symmetry   can  be broken    by  generic
background gauge curvature  field $B_{\mu\nu}$. More particularly, the
string worldsheet  will    contain  an axial  current  with   internal
components   $j^{5i}  $ $=  \overline  \Psi \mbox{\boldmath$\gamma$}^5
\mbox{\boldmath$\gamma$}^i\Psi$ that  would   acquires a corresponding
divergence of the  form $ j^{5i}_{\ \  ;i}$ $ \propto \varepsilon^{ij}
G_{ij} \ ,$ where the pull back of the gauge curvature onto the string
worldsheet is defined by $G_{ij}=G_{\mu\nu} x^\mu_{,i} x^\nu_{,j}$. In
practice however,  in the     particular case of     a  Nielsen-Olesen
background,   there   will be no    such   contribution  because in  a
configuration  of this type the   associated gauge curvature will only
have  components in directions orthogonal to  the world sheet, i.e. it
will have vanishing tangentially projected components, $G_{ij}=0$.

Let us now consider what would happen if the  fermions were coupled to
an  independent   long range gauge  field  $A_\mu$  associated with an
electromagnetic type curvature field $F_{\mu\nu} = \partial _\mu A_\nu
-\partial_\nu  A_\mu$ which  can     give rise  to  chiral   anomalies
proportional    to       $\varepsilon^{\mu\nu\rho\sigma}    F_{\mu\nu}
F_{\rho\sigma}$ in the four dimensional theory.

What one finds~\cite{Witten} for fermions having a definite handedness
[like right-handed  neutrinos in the $SO(10)\to  SU(5)$ model] is that
they  contribute  to the $QQR$  and  $QRR$ triangle  anomalies with an
amount that depends on their $Q-$charges in such a way that the theory
is anomaly-free  if  and only   if  $$  \sum  _{_L}  Q^2 =   \sum_{_R}
Q^2,\label{sumQ} $$ where the  sums are taken  over the left and right
handed particles respectively.   It  follows that a theory   involving
only, let us say, left handed particles -- so that the right hand side
of this equation  automatically vanishes -- can  be anomally free only
if the left hand side of this equation also vanishes, i.e. only if all
the  relevant coupling charges   are  zero.   The absence   of  charge
coupling is  therefore essential for  a  theory of  the kind providing
purely chiral fermionic   string currents,  as illustrated in   figure
\ref{fig2}.

\section*{Acknowledgements.} The authors wish to  thank Anne Davis,
Stephen Davis, and Warren Perkins for helpful conversations.

\end{document}